\documentclass[showpacs,aps,prb,reprint,superscriptaddress]{revtex4-2}
\usepackage{bm}
\usepackage{graphicx}
\usepackage{amsmath}
\usepackage{amsfonts}
\usepackage{amssymb}
\usepackage{mathtools}
\usepackage{epstopdf}
\usepackage{hyperref}
\usepackage{braket}
\usepackage{float}
\hypersetup{
    colorlinks,%
    citecolor=blue,%
    linkcolor=blue,%
    urlcolor=blue
}

\begin{document}

    \newcommand{\be}   {\begin{equation}}
    \newcommand{\ee}   {\end{equation}}
    \newcommand{\ba}   {\begin{eqnarray}}
    \newcommand{\ea}   {\end{eqnarray}}
    \newcommand{\ve}  {\varepsilon}

    %%%%%%%%%%%%%%%%%%%%%%%%%%%%%%%%%%%%%%%%%%%%%%%%%%%%%%%%%%%%%%%%%%%%%%%%%%%%%%%
    \title{Quantum dots as parafermion detectors}
        
    \author{Raphael L.~R.~C. Teixeira}
    \affiliation{Instituto de F\'{\i}sica, Universidade de S\~{a}o Paulo,
        C.P.\ 66318, 05315--970 S\~{a}o Paulo, SP, Brazil}
    \author{Luis G.~G.~V. Dias da Silva}
    \affiliation{Instituto de F\'{\i}sica, Universidade de S\~{a}o Paulo,
        C.P.\ 66318, 05315--970 S\~{a}o Paulo, SP, Brazil}
    
    \date{ \today }
    
    \begin{abstract}
    
Parafermionic zero modes, $\mathbb{Z}_n$-symmetric generalizations of the well-known $\mathbb{Z}_2$ Majorana zero modes, can emerge as edge states in topologically nontrivial strongly correlated systems displaying fractionalized excitations. In this paper, we investigate how signatures of  parafermionic zero modes can be detected by its effects on the properties of a quantum dot tunnel-coupled to a system hosting such states. Concretely, we consider a strongly-correlated 1D fermionic model supporting $\mathbb{Z}_4$ parafermionic zero modes coupled to an interacting quantum dot at one of its ends. By using a combination of density matrix renormalization group calculations and analytical  approaches,  we show that the dot's zero-energy spectral function and average occupation numbers can be used to distinguish between  trivial, $\mathbb{Z}_4$ and $2\times \mathbb{Z}_2$ phases of the system. The present work opens the prospect of using quantum dots as detection tools to probe non-trivial topological phases in strongly correlated systems.

    \end{abstract} 
    %\pacs{ APS does not use it anymore}
    %\keywords{Quantum Spin-Hall effect, Edge transport, Topological insulators}   
    \maketitle

    %%%%%%%%%%%%%%%%%%%%%%%%%%%%%%%%%%%%%%%%%%%%%%%%%%%%%%%%%%%%%%%%%%%%%%%%%%%%%%%

\section{Introduction}
\label{sec:Intro}
 
The production and detection of quasiparticles with statistics which are neither fermionic or bosonic is a fundamental quest in condensed matter physics. Proposals for the realization of such quasiparticles (dubbed \textit{anyons} \cite{Wilczek:Phys.Rev.Lett.:957959:1982}) have been put forward over the years, and recent experimental findings seem to confirm their existence \cite{Nakamura:NaturePhysics:931936:2020,Bartolomei:Science:173177:2020}. A particular type of anyons with \textit{non-Abelian} exchange statistics has been gathering attention for the past few years \cite{Clarke:Non-AbelianAyonsFQH:2013} as their exotic properties make them ideal platforms to realize topological quantum computers (TQCs) \cite{Nayak:NonAbelianAnyonTQC:2008,Mong:UniversalTQC:2014,Aasen2016}.

Majorana zero modes (MZMs) are currently the main candidates for realizing TQCs based on non-Abelian anyonic exchange statistics of the Ising type \cite{Kitaev:P.U:2001,Alicea:Reports:2012,Nayak:MajoranaQuantumComputation:2015}. However, quantum gates based on Ising braiding are, by definition, limited in scope. The reason being that the braiding of Ising anyons amounts to a $\pi/2$ qbit rotation in the Block sphere \cite{Nayak:NonAbelianAnyonTQC:2008}. As such, the prospect of using parafermionic modes as the building blocks for more generic quantum gates can expand these possibilities due to their  Fibonacci-type braiding statistics \cite{Alicea:AnnualReviewofCondensedMatterPhysics:119139:2016,Loss:QCwithParafermion:2016}.

Parafermionic modes can be viewed as $\mathbb{Z}_n$ generalizations of $\mathbb{Z}_2$-symmetric MZMs. They were first proposed in the context of clock-models \cite{Fradkin:Parafermions:1980,Fendley:ParafermionicZeroMode:2012}, and later used to describe exotic fractional quantum Hall excitations   \cite{Read:ParafermionIncompressibleStates:1999}. Recently, parafermions have been subjected to renewed interest \cite{Stoudenmire:AssemblingFibonacci:2015,Groenendijk:ParafermionBraidingFQH:2019,Mazza:AnyonicTBofparafermions:2019,Oreg:SignaturesParafermionFQHE:2020,Klinovaja:CornerSatetBilayerGraphene:2020}, as parallels of parafermionic- and MZM-hosting systems were suggested \cite{Fendley:ParafermionicZeroMode:2012,Fendley:StabilityParafermion:2014}. 

Due to their unusual nature, proposals for the experimental realization of parafermionic zero modes (PZMs) usually rely on finding $\mathbb{Z}_n$-symmetric ground states of  effective low-energy models \cite{Klinovaja:ParafermionsNanowireBundle:2014,Klinovaja:TRIParafermionRashba:2014,Fleckenstein:ParafermionHEdgeQSHI:2018}. Only recently a Kitaev-type lattice model hosting free PZMs was mapped into a strongly interacting model of (spinful) fermionic particles in a 1D lattice  \cite{Calzona:Z4FermionLattice:2018,Alicea:FermionezParafermionsSEMajorana:2018}. Although these  models might look somewhat unrealistic due to the presence of rather exotic three-body interaction terms, they offer a concrete path to realizations of parafermions in electronic systems, similar to the role the Kitaev chain played for the  Majorana zero modes almost 20 years ago \cite{Kitaev:P.U:2001}. Nonetheless, several questions remain open, from possible realizations of different parafermions to their experimental signature.

 In this paper, we address these questions by proposing the use of quantum dots (QD) as an experimental probe to detect the signature of parafermionic modes similar to zero-bias peaks predicted in Majorana-quantum dot setups \cite{Leijnse:MeasuringMajQDLife:2011,Liu:MajoranaQD:2011,Vernek:MajoarnaLeakage:2014,David:InteractionsMajoranaQD:2015,Prada:MajoranaQD:2017}. Here, we focus on QDs coupled to topological 1D systems hosting  $\mathbb{Z}_4$ PZMs at their edges \cite{Zhang:TRIZ4FractionJoesephsonEffect,vonOppen:Z4parafermionQSHJosephsonImp:2017,Alicea:FermionezParafermionsSEMajorana:2018,Calzona:Z4FermionLattice:2018}.

 Our results show that experimentally readily accessible QD properties such as the local density of states can be used to distinguish the different topological phases of the system, indicating the presence or absence of edge PZMs. More importantly, the QD signatures can distinguish between phases of local $\mathbb{Z}_4$ parafermionic modes and those comprised of two $\mathbb{Z}_2$ Majorana modes.

The paper is organized as follows: in Sec.~\ref{sec:model}, we introduce the model Hamiltonian for a chain with dangling parafermion modes and the coupling term to an interacting quantum dot. The system's phase diagram, calculated with DMRG, is presented in Sec.~\ref{sec:phases}, along with results showing that  zero-energy density of states calculated at the dot site can probe the different phases. These results are further discussed in Sec.~\ref{sec:detection}, where we show how the dot's LDOS and average occupancy can be used to distinguish between trivial, $\mathbb{Z}_4$ and $2\times \mathbb{Z}_2$ phases. In Sec.~\ref{sec:analytic}, we show that the DMRG results for the $\mathbb{Z}_4$ phase can be understood by an analytical calculation of the first-order corrections in the dot coupling. Finally, our concluding remarks are presented in Sec.~\ref{sec:conclusion}. 

\section{Fermionic model}
\label{sec:model}

\begin{figure}[t]
	\begin{center}
		\includegraphics[width=1\columnwidth]{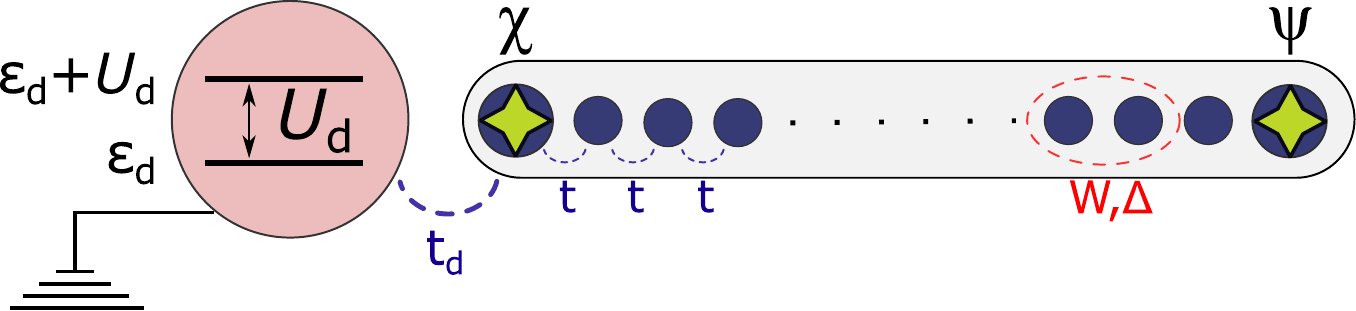}
		\caption{Parafermion chain coupled to a quantum dot. The dot has an electron-electron repulsion given by $U_d$ and an energy given by $\epsilon_d$.} 
		\label{fig:syst}
	\end{center}
\end{figure}

We consider a setup composed of a quantum dot coupled to a 1D fermionic chain that hosts $\mathbb{Z}_4$ parafermionic modes at its ends, as depicted in Fig.~\ref{fig:syst}. The first challenge is to devise a system of correlated 1D spinful fermions which can host such $\mathbb{Z}_4$ parafermionic modes. A promising path is to express parafermionic operators in terms of purely fermionic ones  \cite{Fendley:ParafermionicZeroMode:2012,Fendley:StabilityParafermion:2014} and then write a Kitaev-like model for $\mathbb{Z}_4$ parafermions as a strongly-correlated fermionic model in 1D with local terms only \cite{Calzona:Z4FermionLattice:2018}. Such transformation will generate (nearest neighbor) superconducting and two- and three-body interaction terms.  After collecting these terms, we can write the following Hamiltonian for the model as 
\begin{equation}
\label{eq:z4Hamiltonian}
H_{Z_4}=H_{\rm SC} + H_{W} \; , 
\end{equation}
with
\begin{align}
H_{\rm SC}=& \; - \sum_{\sigma, j} t c_{\sigma, j}^{\dagger} c_{\sigma, j+1}-i \Delta c_{\bar{\sigma}, j}^{\dagger} c_{\sigma, j+1}^{\dagger} \label{eq:HSC} \\
H_{W}=& \; -W \sum_{\sigma, j} \left[c_{\sigma, j}^{\dagger} c_{\sigma, j+1}\left(-n_{\bar{\sigma}, j}-n_{\bar{\sigma}, j+1}\right)\right.\nonumber\\&\hspace{1.2cm} +c_{\sigma, j}^{\dagger} c_{\sigma, j+1}^{\dagger}\left(n_{\bar{\sigma}, j}-n_{\bar{\sigma}, j+1}\right) \nonumber\\ &\hspace{1.2cm}+ i c_{\sigma, j}^{\dagger} c_{\bar{\sigma}, j+1}\left(n_{\bar{\sigma}, j} - n_{\sigma, j+1}\right)^2 \nonumber\\&\hspace{1.2cm}\left.+ i c_{\sigma, j}^{\dagger} c_{\bar{\sigma}, j+1}^{\dagger}\left( n_{\bar{\sigma}, j}- n_{\sigma, j+1}\right )^2
%\left(2 n_{\sigma, j} n_{\bar{\sigma}, j+1}\right )
\right]+ \mbox{H.c.} \; , \label{eq:HW} 
\end{align}
where $t$ is the (single-particle) hopping parameter, $\Delta$ is an unconventional superconductivity order parameter (assumed real) that couples different spins in neighbour sites and $W$ is the strength of 2 and 3-body interactions. The many-body interactions in Eq.~\eqref{eq:HW} have different behaviors and can be seen as a competition in the system that tries to push the ground-state away from the half occupation limit. 

Let us briefly discuss the four interaction terms in Eq.~\eqref{eq:HW} in more detail. The first is essentially a hopping term that is hindered when there are no electrons of opposite spins in the two hopping sites. As such, it can be understood as an effective two-body attraction between the electrons of opposite spins. The second term describes a p-wave superconducting pairing which depends on the two sites having  distinct opposite spin occupation numbers. This, in turn, thwarts the creation of a p-wave pair of a given spin unless there is a charge imbalance of electrons with opposite spin in the two sites. The third and forth terms are, respectively, three-body spin-orbit-like hopping and spin-mixing p-wave paring terms which contribute only when 
two neighboring sites have distinct occupation numbers of opposite spin.

This model has two important features. In the limit $t=\Delta=W\equiv t $, the Hamiltonian maps exactly \cite{Calzona:Z4FermionLattice:2018} into a Kitaev-like chain of $\mathbb{Z}_4$ parafermions with two uncoupled parafermions at its ends, namely: 
\begin{equation}
H_{\rm pf} = -J e^{-i\pi/4} \sum_{j=1}^{L-1}\psi_j\chi_{j+1}^\dagger + \mbox{H.c.} \;.
\label{eq:HamiltonianPFz}
\end{equation}
where $\chi$ and $\psi$ are $\mathbb{Z}_4$ parafermions satisfying $\chi_j^\dagger=\chi_j^3$,  $\psi_j^\dagger=\psi_j^3$ and $\chi_j \chi_k = i \chi_k \chi_j $, $\psi_j \psi_k = i \psi_k \psi_j $ $\chi_j \chi_k = i \chi_k \chi_j $ for $j<k$ and $\chi_j \psi_k = i \psi_k \chi_j $  for $j\leq k$. At the same time, the limit $t=\Delta$ with $W=0$ gives a chain with two Majorana modes at each end ($2\times \mathbb{Z}_2$) \cite{Calzona:Z4FermionLattice:2018}. As such, we can explore trivial, $\mathbb{Z}_4$ and $2\times \mathbb{Z}_2$ phases just by varying $\Delta$ and $W$.

We consider the case there the chain is coupled to an interacting  quantum dot located at it's left end, as depicted in Fig.~\ref{fig:syst}. The Hamiltonian of the full system is 
\begin{equation}
\label{eq:Z4_QD}
H_{\rm Z_4-QD}=H_{Z_4}+H_{\rm QD}+H_{\rm pf-QD} \; ,
\end{equation}
where
\begin{align}
\label{eq:coupHop}
H_{\rm pf-QD} = & -t_d \sum_{\sigma=\uparrow,\downarrow} c^\dagger_{\sigma,d} c_{\sigma,1}-c^\dagger_{\sigma,d}c^\dagger_{\sigma,1}+\mbox{H.c.} \; , \\
H_{\rm QD} = & U_d n_{\uparrow,d}n_{\downarrow,d}+ \epsilon_d(n_{\uparrow,d} + n_{\uparrow,d})\label{eq:HamQD} \; .%\mbox{H.c.}
\end{align}

In the above, $c^\dagger_{\sigma,d}$ ($c_{\sigma,d}$) represents a creation (destruction) operator for an electron of spin $\sigma$ in the dot with  $n_{\sigma,d} \equiv c^\dagger_{\sigma,d}c_{\sigma,d}$. $H_{\rm pf-QD}$ in Eq.\ \eqref{eq:coupHop} represents the dot-chain coupling. Notice that it includes an Andreev-reflection term \footnote{The choice of a minus sign in the Andreev-like coupling in  Eq.\ \eqref{eq:coupHop}  does not affect the results and any phase factor $e^{i\theta}$ would work, except $\theta=0$ (plus sign). In this case we have an transition similar to what happens with Majoranas, but shifted by $\pi$.}, similarly to the case of quantum dots coupled to chains hosting MZMs \cite{Vernek:MajoarnaLeakage:2014,Prada:MajoranaQD:2017}. In addition, the quantum dot Hamiltonian is given by Eq.\ \eqref{eq:HamQD}, which contains an electron-electron repulsion term with strength by $U_d$ and a (tunable) single-particle energy level at $\epsilon_d$. With no loss of generality, we take $t_d=0.1 t$ throughout the paper.

\section{Phase diagram}
\label{sec:phases}

The phase diagram of the system can be obtaining by following the many-body ground state degeneracies as well as the gap to the first excited states of either the chain-only or chain+quantum dot systems. We obtain the overall ground-states of the respective Hamiltonians  with the DMRG method \cite{White:Phys.Rev.B:10345:1993,Schollwock:DMRG:2005,Schollwock:DMRG-MPS:2011} as implemented within the ITensor package \cite{ITensor}. 

Ground-state degeneracy count plays an important role in distinguishing the two topological phases from the trivial one: while the ground state is four-fold degenerate in the first two, it is always non-degenerate in the latter.  To this end, we determine the degeneracy of the ground state by counting the number of low-lying states within a window $\delta E \lesssim 10^{-6} t$. This value is well within the ground-state energy  accuracy in the DMRG calculations given the bond dimension and chain lengths used (see Appendix \ref{Sec:FiniteSize} for more details). It is also enough to characterize gap openings between the ground state and the first excited state, which, for the parameters used, are of order $\sim 10^{-3} t$ in the trivial phase and $\gtrsim 0.1 t$ in the topological phases.

\begin{figure}[t]
	\begin{center}
		\includegraphics[width=0.99\columnwidth]{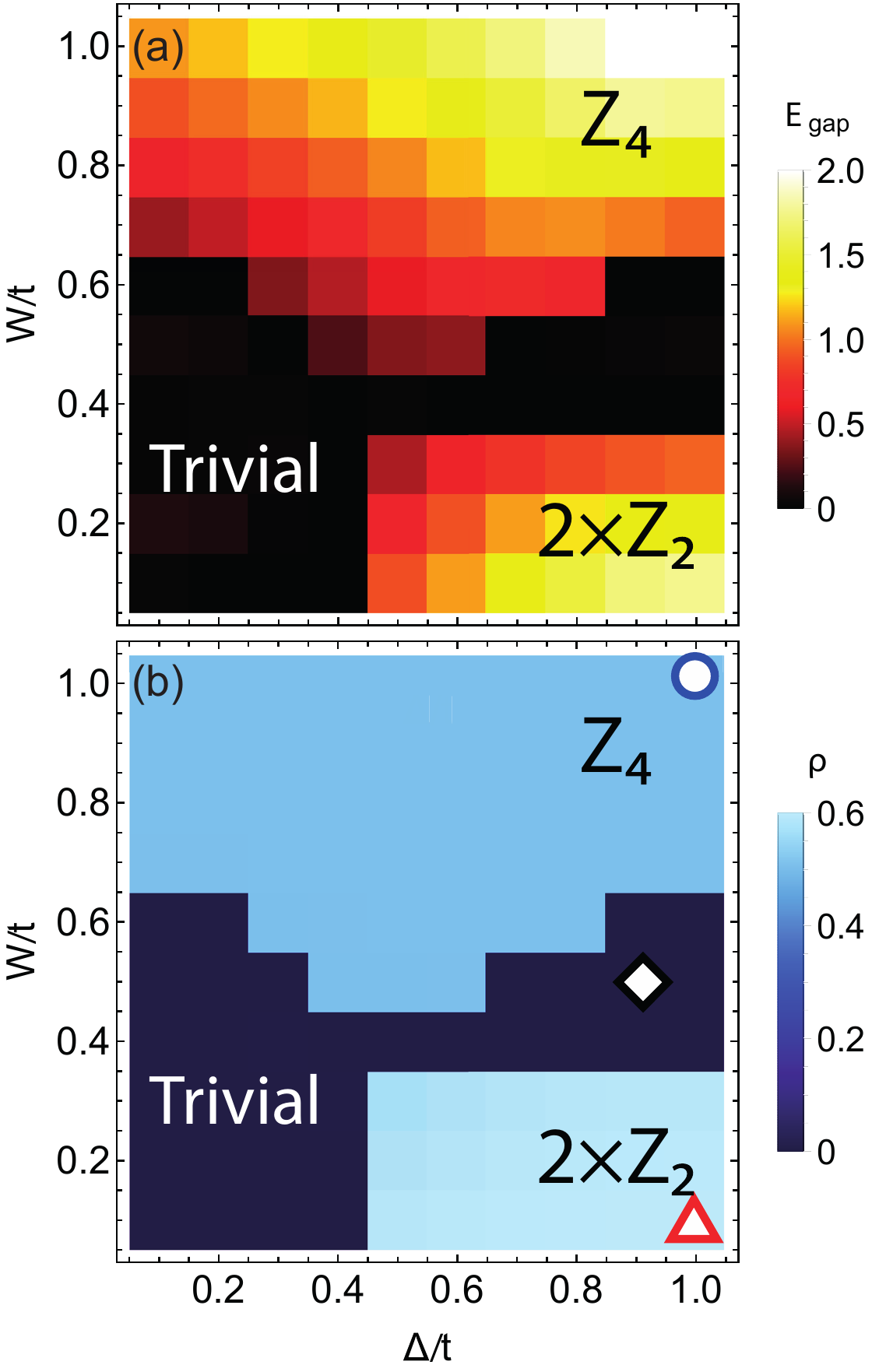}
		\caption{Phase diagrams of $H_{Z_4}$, Eq.\ \eqref{eq:HW}. (a) Energy gap between ground and first excited states for a 20-site chain described by $H_{Z_4}$. (b) Quantum dot LDOS $\rho_d(0)/(2\pi)$ for a 20-site chain attached to the QD for $U_d/t=1$ and $\epsilon_d=0$. Symbols represent the $\Delta$ and $W$ values used in the curves shown in Fig.~\ref{fig:Z4x2Z2}.
		} %
		\label{fig:phases}
	\end{center}
\end{figure}

In the chain-only case, the topological phases of $H_{Z_4}$ were obtained by computing the gap between ground and first excited states of a $20$-site chain. These are shown in Fig.~\ref{fig:phases}(a) for different values of the parameters $W$ and $\Delta$. Analytical solutions exist for three out of the four corners of the phase diagram, namely $\Delta/t=W/t=1$, $\Delta/t=1$, $W=0$ and $\Delta=W=0$. Those limits correspond respectively to topological phases $\mathbb{Z}_4$, $2\times \mathbb{Z}_2$ and ``trivial'', i.e. a simple tight-binding chain. As $\Delta$ and $W$ are varied, topological phase transitions occur as the gap goes to zero. By following these gap closings and comparing with the analytical limits, we can determine which region corresponds to each phase. 

We stress that many-body interactions play an essential role in the transition to the $\mathbb{Z}_4$ parafermion phase. In fact, as can be seen in  Fig.~\ref{fig:phases}(a), the emergence of $\mathbb{Z}_4$ parafermionic modes occurs only for $W/t>0.4$. Concurrently, $2\times \mathbb{Z}_2$ Majorana phase occurs for weak many-body interaction and a large values of the superconducting order parameter $\Delta$.

\section{Parafermion detection}
\label{sec:detection}

Detecting topological phase transitions by monitoring the gap and ground state degeneracies can be a challenging task. Not only it is difficult to tell the $\mathbb{Z}_4$ and $2 \times \mathbb{Z}_2$ topological phases from each other but also  finite-size effects  can be an issue, as discussed in Appendix~\ref{Sec:FiniteSize}.
Interestingly, we find that these phases can be also be probed by accessing the local density of states of a quantum dot side-coupled to the system.  The dot's occupation can also be used to differentiate the phases, making the dot an ideal platform to detect parafermions. Together, these features can give a clearer experimental signature of the topological phase transitions in the system. 

More importantly, our results establish a one-to-one correspondence between the zero-energy density of states and the  different topological and non-topological phases, allowing for a clear signature of the presence or absence of PZMs in the chain. 
This correspondence is nicely illustrated by comparing Figs.~\ref{fig:phases}(a) and (b) and constitute one of the main results of this work.

\subsection{Local density of states}
\label{sec:LDOS}

The local density of states (LDOS) for a given site in the chain can be accessed by tracking the matrix elements of the local fermionic operators between the $N_{\rm gs}$ ground states of the system \cite{Mazza:ParafermionGS:2017,Calzona:Z4FermionLattice:2018}. We follow this route to obtain the QD LDOS from the zero-energy spectral function given by:
\begin{equation}\label{eq:rho}
\rho_d(0) = \frac{2\pi}{N_{\rm gs}} \sum_{\sigma,\ket{g},\ket{g^\prime}}|\bra{g}c^\dagger_{\sigma,d}\ket{g^\prime}|^2+|\bra{g}c_{\sigma,d}\ket{g^\prime}|^2 \; ,
\end{equation}
where we sum over all $N_{\rm gs}$ ground states $\ket{g}$,$\ket{g^\prime}$ of $H_{\rm Z_4-QD}$ (Eq.~\eqref{eq:Z4_QD}). In practice, the sum in Eq.~\eqref{eq:rho} is comprised of $N_{\rm gs}$ identical terms. Thus, it is sufficient to calculate only one of these terms for a given ``reference'' ground state $\ket{g^\prime} \equiv \ket{0}$, which we choose as the first state with the lowest energy computed by DMRG.  

\begin{figure}
	\begin{center}
		\includegraphics[width=0.99\columnwidth]{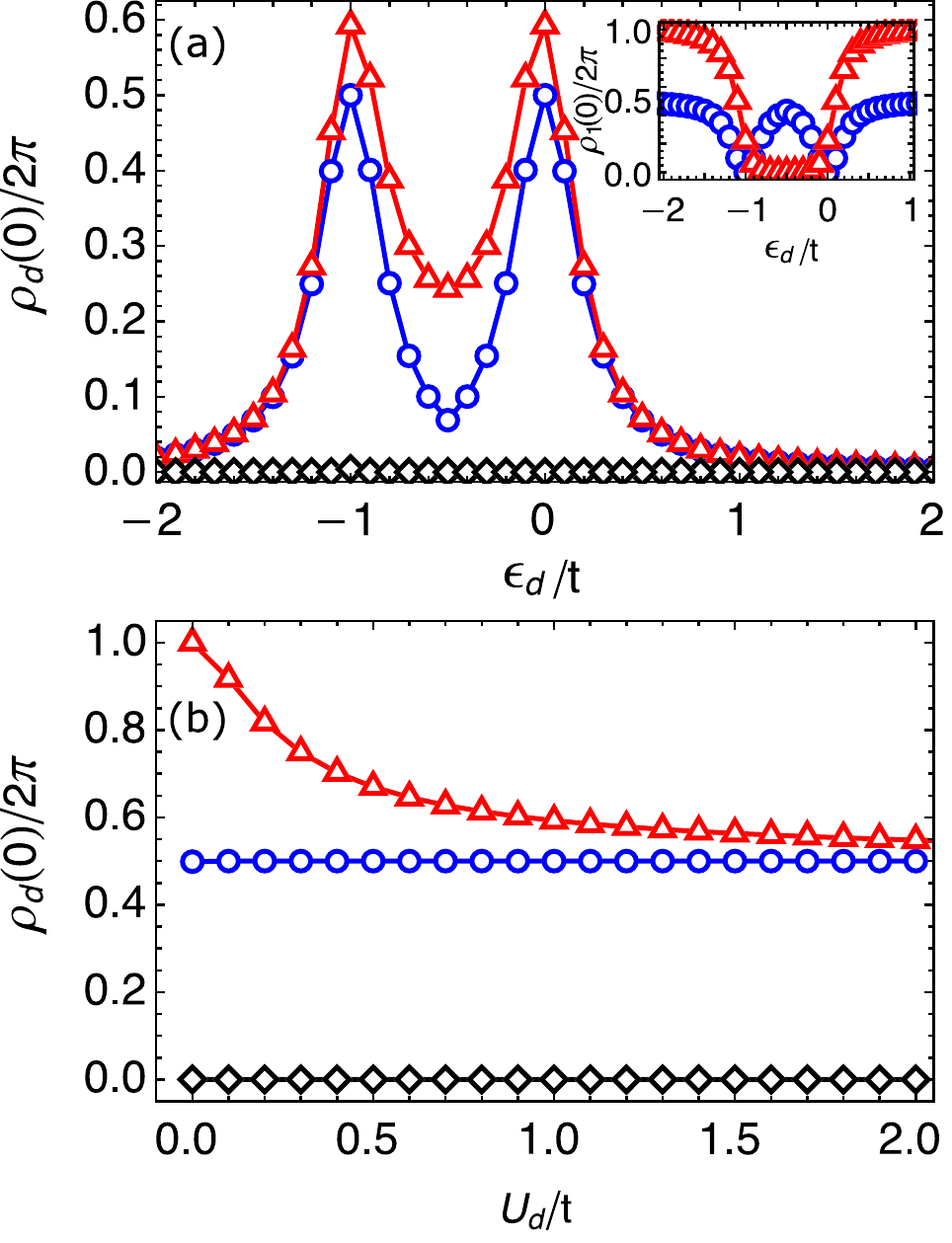}
		\caption{(a) QD LDOS versus $\epsilon_d$ calculated for  $U_d/t=1$ and chain  parameters corresponding to the symbols marked in Fig.~\ref{fig:phases}(b): $W/t=\Delta/t=1$ (blue circles, $\mathbb{Z}_4$ phase) ;  $\Delta/t=1$, $W=0$ (red triangles, $2\times \mathbb{Z}_2$ phase); $W/t=0.5$, $\Delta/t=0.9$, (black diamonds, trivial phase). The inset shows the LDOS at the first site of the chain for the same parameters. (b) LDOS vs electron-electron interaction $U_d$ at $\epsilon_d=0$. } 
		\label{fig:Z4x2Z2}
	\end{center}
\end{figure}

We can compare the phase diagram due the gap to the phase diagram due the dot's zero-energy DOS, Fig,~\ref{fig:phases}(b). The LDOS phase diagram was obtained  for $U_d/t=1$, $\epsilon_d=0$ and   $\rho_d(0)$ assumes a characteristic, near constant, non-zero value at each of the topological phases while it drops to zero in the transition to the trivial phase.

The characteristic values of $\rho_d(0)$ on each topological phase depend on $U_d$ and $\epsilon_d$, as shown in Fig.~\ref{fig:Z4x2Z2}.  As a general feature, $\rho_d(0)$ displays peaks at $\epsilon_d=0$ and $\epsilon_d=-U_d$, as shown in Fig.~\ref{fig:Z4x2Z2}(a) in the topological phases. Generically, $\rho_d(0)$ can distinguish the different phases by gate-tuning the quantum dot to the single-occupation regime $-U_d < \epsilon_d < 0$. In fact, tuning the dot to the particle-hole symmetric point $\epsilon_d = -U_d/2$ can maximize its sensibility to distinguish the different phases. Here, the $\rho_d(0)$ value at the $2\times \mathbb{Z}_2$ is nearly twice that of the value at the $\mathbb{Z}_4$ phase.

The values of $\rho_d(0)$ at the peaks can be used to differentiate the $\mathbb{Z}_4$ and $2\times \mathbb{Z}_2$ phases. While the $\mathbb{Z}_4$ phase has a value of $\rho_d(0)/2\pi \sim 0.5$  at the peaks the $2\times \mathbb{Z}_2$ phase has a larger value $\rho_d(0)\sim 0.58$ for $U_d/t=1$. These values are a consequence of the strong localization of the ground state in both phases (at least half of the total spectral weight) at the QD site. This situation is similar to the ``leaking'' of Majorana bound states into quantum dots studied in Refs. \cite{Vernek:MajoarnaLeakage:2014,David:InteractionsMajoranaQD:2015}.

The ``leaking'' is stronger for the case of MZMs ($2 \times \mathbb{Z}_2$ phase) than for PZMs ($\mathbb{Z}_4$ phase). This is illustrated in the inset of Fig.~\ref{fig:Z4x2Z2}(a) which shows the LDOS at the first site of the chain $\rho_1(0)$. For the $\mathbb{Z}_4$ phase, we find $\rho_1(0)=\pi-\rho_d(0)$, reaching $\rho_1(0)  \approx \pi$ and $\rho_d(0)  \approx 0$ (localized in the chain rather than in the dot) for  $\epsilon_d = -U_d/2$ and $\epsilon_d >0$, $\epsilon_d < -U_d$. This indicates that the PZM ``leaks'' into the dot only at the Coulomb peaks $\epsilon_d=0 , -U_d$. In the  $2\times \mathbb{Z}_2$ phase, by contrast, $\rho_1(0)\sim0$ for $-U_d<\epsilon_d<0$, implying a much stronger leaking of the two MZMs into the dot.

Moreover, the  $\rho_d(0)$ value in the $\mathbb{Z}_4$ phase is essentially independent of the  electron-electron interaction in the dot $U_d$, as shown in Fig.~\ref{fig:Z4x2Z2}(b) for $\epsilon_d =0$ . By contrast, increasing values of $U_d$ tend to decrease the $\rho_d(0)$ value at the $2\times \mathbb{Z}_2$ $\rho_d(0)$. This indicates that QDs with $U_d \sim t$ can be more efficient in distinguishing the different topological phases.

\subsection{Dot occupation}
\label{sec:ndot}

\begin{figure}[t]
	\begin{center}
		\includegraphics[width=0.99\columnwidth]{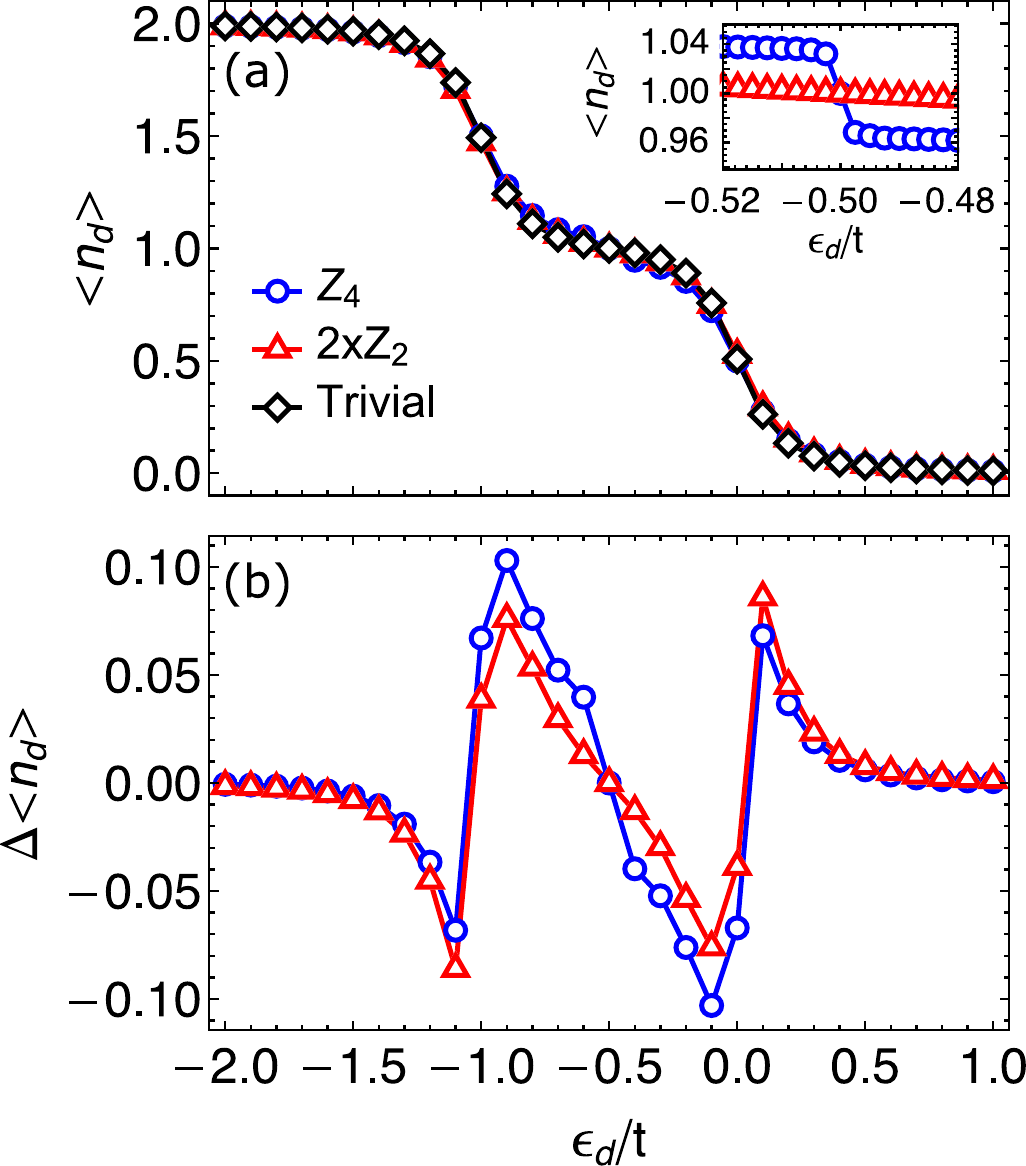}
		\caption{ (a) QD occupancy $\langle n_d \rangle$ vs $\epsilon_d$ for the same parameters as in Fig.~\ref{fig:Z4x2Z2}. Inset: enhancement showing a discontinuity  in $\langle n_d \rangle$ calculated at the $\mathbb{Z}_4$ phase at $\epsilon_d=-U/2$. (b) Occupancy difference between topological and trivial phases.} 
		\label{fig:occ}
	\end{center}
\end{figure}

As discussed above, the stronger signatures of PZMs in the dot LDOS occur precisely at the points where the dot's occupancy changes, either from from unoccupied to singly occupied ($\epsilon_d \approx 0$) as well as from singly occupied to doubly occupied ($\epsilon_d \approx -U_d$). In fact, one can track the presence/absence of PZMs in the chain by monitoring the average occupation of the quantum dot.

This is shown in  Fig.~\ref{fig:occ}(a), where we show the zero-temperature dot occupancy $\langle n_d \rangle$ versus $\epsilon_d$ for each of the phases at $W/t=\Delta/t=1$ ($\mathbb{Z}_4$), $W/t=0$ $\Delta/t=1$ ($2\times \mathbb{Z}_2$), $W/t=0.5$ $\Delta/t=0.9$ (trivial), shown in the phase diagram. Although the overall behavior of the occupancy is similar, with well-defined occupancy plateaus as a function of $\epsilon_d$, there are subtle differences depending on the phase of the system. 

For instance, while both trivial and $2\times \mathbb{Z}_2$ phase display a smooth change in occupation number around the symmetric point $\epsilon_d=-U_d/2$,  in the $\mathbb{Z}_4$ phase  the occupancy jumps from around $0.96$ at $\epsilon_d>-U/2$ to exactly 1 at $-U/2$ than to $1.04$ at $\epsilon_d<-U/2$, (inset of Fig.~\ref{fig:occ}(a)). \footnote{We verified that similar results were obtained for more generic parameters}. In all cases, we confirmed that there is no spin-polarization in the occupancy (namely $\langle n_{d \uparrow} \rangle = \langle n_{d \downarrow} \rangle$). 

The distinction between the curves at the different phases can be better appreciated by subtracting $\langle n_d \rangle(\epsilon_d)$ from the trivial case,  $\Delta \langle n_d \rangle \equiv \langle n_d \rangle - \langle n_d \rangle_{\rm trivial}$, as plotted in  Fig.~\ref{fig:occ}(b). In particular, $\Delta \langle n_d \rangle$ changes  rather strongly near the inflection points $\epsilon_d = 0, -U_d$, allowing one to differentiate the topological phases from trivial one and from each other.

\section{Comparison with analytic results}
\label{sec:analytic}

In order to better understand in DMRG results, we use an analytical perturbative approach to describe the changes in the $\mathbb{Z}_4$ topological phase in the presence of the coupling to the quantum dot.  

Our approximation consists in considering the analytic results for the (four-fold degenerate) ground state $|g^{(0)} \rangle$ of $H_{\rm pf}$ given by Eq.~\eqref{eq:HamiltonianPFz} (which describes the $\mathbb{Z}_4$ phase of $H_{Z_4}$ at $\Delta=W=t$) and calculate the first-order correction due to the coupling $t_d$ to the quantum dot given by Eqs.~\eqref{eq:coupHop}-\eqref{eq:HamQD}. The resulting corrected states $|g^{(1)} \rangle$ are then used in Eq.~\eqref{eq:rho} to obtain an approximation for the dot LDOS $\tilde{\rho}_d(0)$. Details of this procedure are given in   Appendix \ref{Sec:FirstO}.

\begin{figure}
	\begin{center}
		\includegraphics[width=0.99\columnwidth]{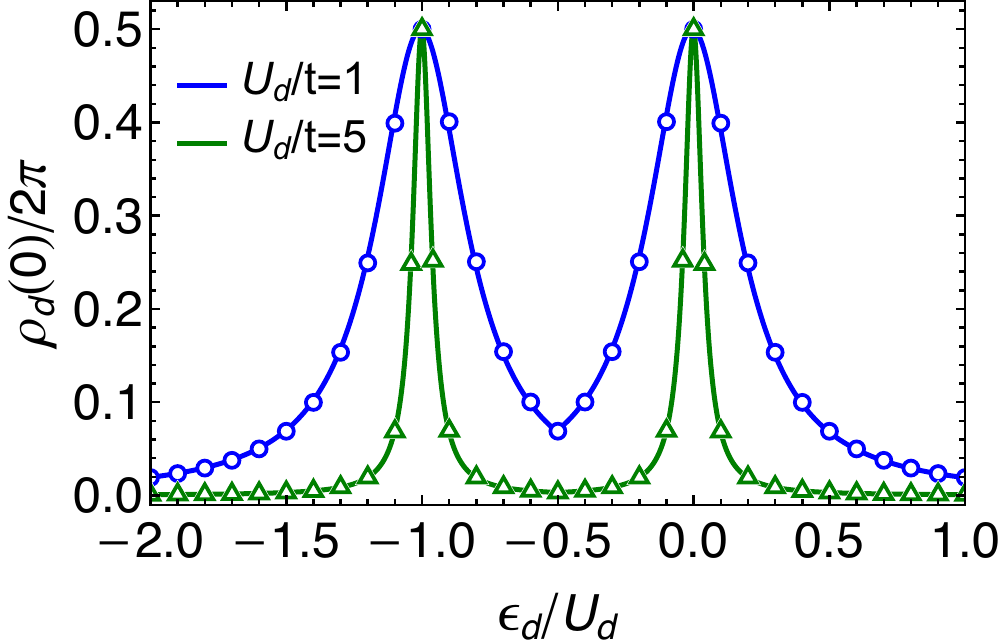}
		\caption{Comparison between the first-order approximation approach (lines) and DMRG results (symbols) for the $\mathbb{Z}_4$ phase with $U_d/t=1$ (blue), and $U_d/t=5$ (green).}
		\label{fig:FirstOrder}
	\end{center}
\end{figure}

One of the artifacts of the approximation is that  $\{ |g^{(1)} \rangle \}$ is now split into two doublets of Fock parafermion dot sates, with an energy splitting of order $ \sim t_d/t$ (see  Appendix \ref{Sec:FirstO}).
Nonetheless, by considering the the lowest energy doublet and calculating the dot LDOS from Eq.~\eqref{eq:rho}, one obtains an excellent agreement with the DMRG calculations, as shown in Fig.~\ref{fig:FirstOrder}.

The LDOS calculated within the analytic approximation can shed some light on the distinct signatures of the presence of PZMs, namely the peaks at $\epsilon_d=0,-U_d$. By closely looking at the perturbed ground state doublet we find that both Fock parafermion states have the same components precisely for $\epsilon_d=0,-U_d$. This matches what one expects for a PZM localized in the dot: an equal-weight linear combination of Fock parafermion states.

\section{Concluding remarks} 
\label{sec:conclusion}

In this work, we propose that quantum dots can be used to probe the presence of parafermionic zero modes in strongly correlated topological systems. Local measurements of quantum dot properties such as the local density of states or the dot's occupancy can discern trivial from topological phases and even tell different topological phases apart from each other. 

We illustrate this by considering a model of a quantum dot coupled to strongly correlated 1D model with a topological phase displaying $\mathbb{Z}_4$ parafermionic edge zero modes. Our DMRG calculations show that the QD properties can map the phase diagram of the topological system in a one-to-one correspondence with the phase diagram obtained by tracking the ground state degeneracy and the opening and closing of energy gaps. In fact, using the QD as a probe has a clear advantage in discerning  $\mathbb{Z}_4$ and $2\times \mathbb{Z}_2$ phases from each other, as they share general features in terms of ground state degeneracy and gap sizes.

The peak height in the QD LDOS as a function of the QD onsite energy $\epsilon_d$ can be used to distinguish the two topological phases from each other and from the trivial one. The main mechanism leading to the LDOS peaks is the ``leaking'' of edge PZMs from the chain to the QD. This leaking is stronger for the $2 \times \mathbb{Z}_2$ phase, which resembles the case o QD-Majorana coupled systems \cite{Vernek:MajoarnaLeakage:2014,David:InteractionsMajoranaQD:2015} and allows a clear distinction of the $\mathbb{Z}_4$ phase, which, in turn shows a strong pinning of the QD LDOS value.

In order to understand better the QD signatures of the $\mathbb{Z}_4$ phase, we calculated the first order correction to the topological ground state due to the coupling to the QD. These analytical results nicely match the DMRG numerics and confirm the presence of a true parafermionic  state localized in the QD site for $\epsilon_d$ values corresponding to the peaks in the LDOS.  

Moreover, the dot charge occupancy $\langle n_d \rangle$ as a function of $\epsilon_d$  can also be used to differentiate the different topological phases in the system. Not only the two topological phases have distinct $\langle n_d \rangle$ vs $\epsilon_d$ curves from the trivial one but the $\mathbb{Z}_4$ phase shows a a discontinuity around $\epsilon_d = -U_d/2$, which does is not present in the trivial and $2\times \mathbb{Z}_2$ phases.

These results indicate that quantum dots can be effectively used as parafermion detectors. Their ability to distinguish between the different  phases, together with the relatively direct experimental access to the dot's local properties, bring interesting prospects in the use of QDs as a tool in the search of parafermionic zero modes.

\acknowledgments

We acknowledge financial support from Brazilian agencies FAPESP (Grant no.  2019/11550-8), Capes, and CNPq (Graduate scholarship program 141556/2018-8, and Research Grants 308351/2017-7, 423137/2018-2, and 309789/2020-6).

 \appendix
 
 \section{Finite-size effects in the phase diagram.\label{Sec:FiniteSize}}

 \begin{figure}[t]
	\begin{center}
		\includegraphics[width=1\columnwidth]{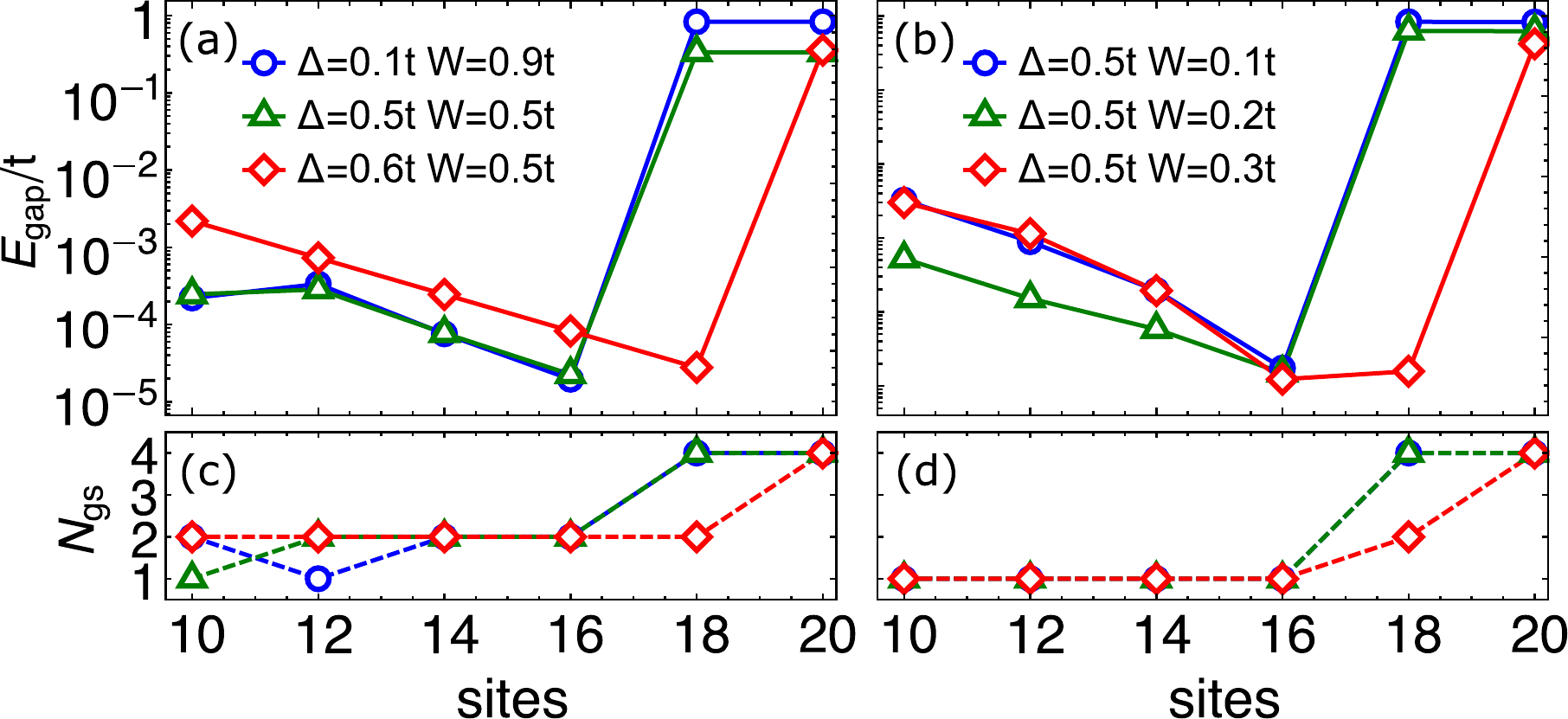}
		\caption{Finite size effects are more prominent in the system without quantum dots. The exponential decay of $E_{gap}$ with the number of sites of the chain, depends not only on the phase (a) $\mathbb{Z}_4$ and (b) $2\times \mathbb{Z}_2$, but also on the values of $\Delta$ and $W$.} 
		\label{fig:LengthZ4}
	\end{center}
\end{figure}

As discussed in the main text, determining whether two states are  ``degenerate'' in the topological phases is an important aspect in constructing the phase diagram show in Fig.~\ref{fig:phases}. The DMRG calculations were carried out using bond dimensions up to 100 (a value usually reached in the trivial phases only) and at least 15 sweeps to ensure convergence. A ``noise term'' was also used to improve convergence to the ground state, avoiding local minima. 

Within these parameters, convergence was obtained within an energy accuracy of $\sim 10^{-8} t, 10^{-9} t$ within the topological phases, which justifies the criteria for considering two states to be degenerate if their energy difference is less than $10^{-6} t$. The energy gap was calculated with similar accuracy by targeting the first few excited states within the same block (no symmetries were considered in the calculations). 

Although such energy gaps can be used to distinguish the topological phase transitions between trivial, $\mathbb{Z}_4$ and $2\times \mathbb{Z}_2$ phases, some care must be taken regarding the system's size used.  For small system sizes, the calculated ``gap'' might have more to to with the overlapping of the edge modes than with the actual ``topological'' gap. This is a similar to the famed ``gap oscillations''  in Majorana systems \cite{DasSarma:Phys.Rev.B:220506:2012}. 

For instance, for $\epsilon_d=0$, zero-energy states tends to localize at the dot site. This can be easily verified for the $\mathbb{Z}_4$ states where the sum of LDOS is constant: for half of the system  $\rho/2\pi=0.5$, and tends to be localized at the dot. Even though the sum of LDOS is not constant in the $2\times \mathbb{Z}_2$ phase, this case also have localization, as we see the decrease of LDOS around half the chain close to the dot.

To illustrate this point, we consider an uncoupled chain described by the Hamiltonian in Eq.~\eqref{eq:z4Hamiltonian}. The dependence of the gap to chain's size is shown in Fig.~\ref{fig:LengthZ4}. An exponential decay in the gap, similar to that predicted for Majorana bound states \cite{DasSarma:Phys.Rev.B:220506:2012}, appear in both topological phases. 

The decay rate with system size at each phase is non-universal and depends on the model's parameters. In Fig.~\ref{fig:LengthZ4}(a) ($\mathbb{Z}_4$ phase),  there are clearly two behaviors, with the gap closing at different rates for $\Delta=0.6$ and $\Delta \leq 0.5$. 
Small deviations from a pure exponential decay are also present, particularly in the $2\times \mathbb{Z}_2$ phase (Fig.~\ref{fig:LengthZ4}(b)). These are probably associated with the details on how the  $2\times \mathbb{Z}_2$ edge states spread along the chain and overlap  with each other. Additionally, some of the ``gap closings'' are, in fact, the formation of a doublet, as illustrated if Fig.~\ref{fig:LengthZ4}(b) for $\Delta=0.5t$ and $W=0.3t$: between  $N=16$ and $N=18$, the ground state degeneracy goes from 1 to 2.

 \section{First-order approximation.\label{Sec:FirstO}}

 In this appendix, we provide an analytical approach to calculate the first-order correction to the ground state of the $\mathbb{Z}_4$ phase in the presence of the quantum dot. 
 
 The starting point is writing the (four-fold degenerate) ground state $\ket{g_j^{(0)}}$ of $H_{\rm pf}$ as $\mathbb{Z}_4$ Fock-parafermion (FPF) states  $\{ |\ket{j}_{pf} \}$, where $j$ is the total FPF number ranging from 0 to 3 \cite{Cobanera:FockParafermion:2014,Mazza:ParafermionGS:2017}.  Notice that one can always write these FPF number basis states in terms of (spinful) fermionic operators acting on a vacuum  state $\ket{0}$, which corresponds to $k\!=\!0$ FPFs \cite{Calzona:Z4FermionLattice:2018}. For instance, for the QD FPF states, we choose $\ket{k\!=\!1}_d=c^\dagger_{\uparrow,d} \ket{0}_d,\ket{k\!=\!2}_d = i c^\dagger_{\uparrow,d} c^\dagger_{\downarrow,d} \ket{0}_d ,\ket{k\!=\!3}_d=-i c^\dagger_{\downarrow, d}\ket{0}_d$.

 Next, we construct a basis for  $H^{(0)} \equiv H_{\rm pf} + H_{\rm QD}$ in the form 
 $\ket{k}_d\otimes\ket{g_j^{(0)}}$ where $\ket{k}_d$ are Fock parafermion states with FPF number $k$ acting on the QD Hilbert space. This gives 16-state basis given by $\ket{k}_d\otimes\ket{g_j^{(0)}}$, where $k$ and $j$ are the total FPF number ranging from 0 to 3 each. To simplify the notation we call $\ket{k}_d\otimes\ket{g_j^{(0)}} \equiv \ket{k,j}$.

The ground state $\ket{g_j^{(0)}}$ of a $L$ site chain is written as a single FPF $\ket{f_a}$ together with a $L-1$ site chain $\ket{s_{j-a}^{(L-1)}}$ with total FPF number $j-a$ mod 4.
\begin{align}\label{eq:PFgs}
\ket{g_j^{(0)}}=\frac{1}{2} &\left( \ket{f_0}\otimes\ket{s_{j}^{(L-1)}}+ \ket{f_1}\otimes\ket{s_{j-1}^{(L-1)}}+\right.\nonumber\\ &\left.+\ket{f_2}\otimes\ket{s_{j-2}^{(L-1)}}+ \ket{f_3}\otimes\ket{s_{j-3}^{(L-1)}}\right).
\end{align}

We also use as a general notation $\ket{n+(k-m)}_d=d^{\dagger \ n}_d d^m_d\ket{k}_d$ where $d(d^\dagger)$ is the annihilation (creation) FPF operator that lowers (rises) the FPF number by one \footnote{In terms of fermionic operators they can be written as \cite{Calzona:Z4FermionLattice:2018} $d_l=i^{\sum_{p<l} (n_{\downarrow,p}+3n_{\uparrow,p}-2n_{\uparrow,p}n_{\downarrow,p})}  c_{\uparrow,l}(1-n_{\downarrow,l}) -  c^\dagger_{\uparrow,l}n_{\downarrow,l} + i c^\dagger_{\downarrow,l}n_{\uparrow,l}$}. If $k-m<0$ or $n+k-m>3$, this state should be understood as zero. 
For a parafermion chain the condition $k-m<0$ or $n+k-m>3$ is not valid, since the ground state with total FPF number $j$ is a sum of all FPF states at the first site, Eq.~\eqref{eq:PFgs}. Instead, we have a filter function  $\eta_n=(4-n)/4$ that arise from applying a FPF operator at the first site of the parafermion chain
\begin{equation}
\bra{g_{n+(j-m)}^{(0)}} d^{\dagger \ n}_1 d^m_1\ket{g_j^{(0)}}=\frac{4-\text{max}(n,m)}{4}.
\end{equation}
Now, we consider fermionic operators in terms of Fock-parafermions at a given site $l$:
\begin{align}
c_{\uparrow, l} =& i^{\sum_{p<l} -N_p +2n_{\uparrow,p}+2n_{\downarrow,p}}\times\nonumber\\
&\times\left(d_l-d_l^\dagger d^2_l - (-1)^{\sum_{p<l} N_p} d^{\dagger 3}_l d^2_l\right),\\
\nonumber\\
c_{\downarrow, l} =& i^{\sum_{p<l} -N_p +2n_{\uparrow,p}+2n_{\downarrow,p}}\times\nonumber\\
&\times(-i)\left((-1)^{\sum_{p<l} N_p} d^{3}_l +d^\dagger_l d_l^2 -d^{\dagger 2}_l d_l^3\right) \; ,
\end{align}
where $N_p$ is the FPF number operator. Notice the string-like phases appearing in the fermionic operators, which is zero for dot operators ($l\!=\!0$).
To simplify the notation,  the string-phase resulting from $c_{\sigma, 1} \ket{k,j}$ (which depends on the dot occupation and the FPF number) is denoted as $\varphi_k$ with $\varphi_0\!=\!1,\varphi_1\!=\!i,\varphi_2\!=\!-1$ and $\varphi_3\!=\!-i$.

 \begin{figure}[t]
	\begin{center}
		\includegraphics[width=1\columnwidth]{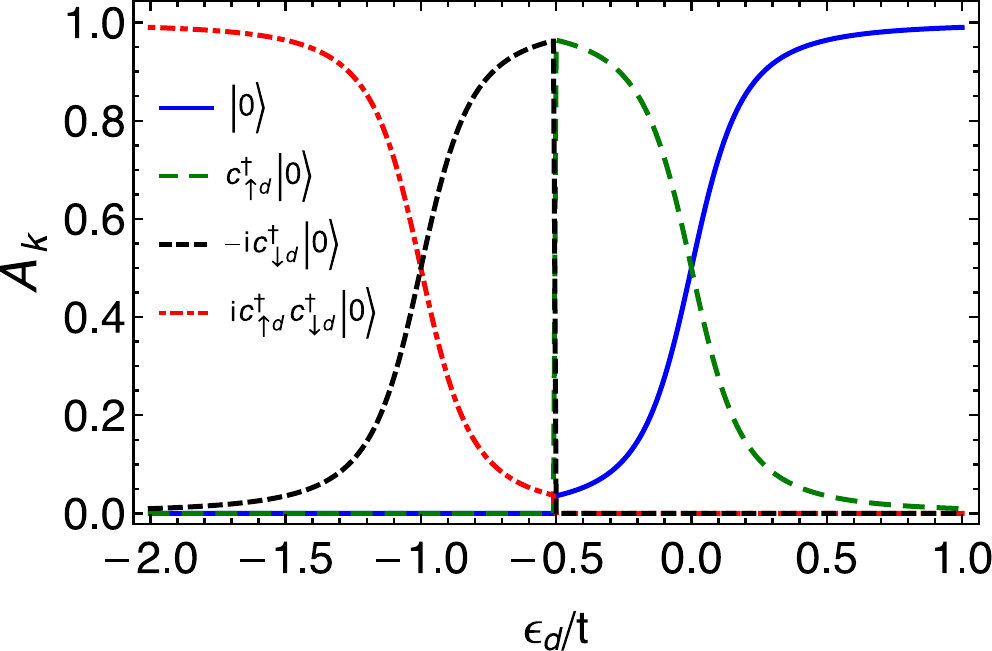}
		\caption{Calculated ground state components  $A_k \equiv \sum_j |\langle k,j | g^{(1)} \rangle|^2$ for each state of the lowest energy doublet in $\ket{g^{(1)}}$ at the dot site. The crossings at $\epsilon_d=-U_d,0$ mark the points where the ground state has equal weights of two FPF states, indicating PZMs localized in the dot. At the symmetric point, $\epsilon_d=-U_d/2$, the ground state doublet changes, resulting in a discontinuity in $A_k$.} 
		\label{fig:spectralWeight}
	\end{center}
\end{figure}

 The next step is to consider the correction to the coupling to the quantum dot $H^{(1)} \equiv H_{\rm pf-QD}$ given by Eqs.~\eqref{eq:coupHop} by calculating its matrix elements in the FPF basis $\{\ket{k,j}\}$. After some straightforward algebra, we can derive the Hamiltonian elements we need, namely:
\begin{align}
c&^\dagger_{d \uparrow} c^\dagger_{1\uparrow}\ket{k,j}=\varphi_k \left[\eta_1\ket{k\!+\!1,\!j\!-\!1}\!-\!\eta_2\ket{k\!+\!1,\!1\!+\!(j\!-\!2)}\!-\right.\nonumber\\
&\left.\!-\varphi^2_k\eta_3\ket{k\!+\!1,\!3\!+(\!j\!-2\!)}  \!-\!\eta_1\ket{2\!+\!(k\!-1\!),\!j\!-\!1}\!+\right.\nonumber\\
&\left.\!+\eta_2\ket{2\!+\!(k\!-\!1)\!,1\!+\!(j\!-\!2)}\!+\!\varphi^2_k\eta_3\ket{2\!+\!(k\!-\!1)\!,3\!+(\!j\!-\!2)}\!-\right.\nonumber\\
&\left.\!-\eta_1\ket{2\!+\!(k\!-\!3),\!j\!-\!1}\!+\!\eta_2\ket{2\!+\!(k\!-\!3),\!1\!+(\!j\!-\!2)}\!+ \right.\nonumber\\
&\left.\!+\varphi^2_k\eta_3\ket{2\!+(\!k\!-\!3),\!3\!+(\!j\!-\!2)} \right] \; ,
\end{align}
\begin{align}
c&^\dagger_{d \downarrow} c_{1\downarrow}\ket{k,j}\!=\varphi_k\! \left[\varphi_k^2\eta_3\ket{k\!+\!3,\!j\!-\!3}\!+\!\eta_2\ket{k\!+\!3,\!1\!+\!(j\!-\!2)}\!-\right.\nonumber\\
&\left.\!-\eta_3\ket{k\!+\!3,\!2\!+\!(j\!-\!3)}\!+\!\varphi_k^2\eta_3\ket{2\!+\!(k\!-\!1)\!,j\!-\!3}\!+  \right.\nonumber\\
&\left.\!+\eta_2\ket{2\!+\!(k\!-\!1),\!1\!+\!(j\!-\!2)}\!-\!\eta_3\ket{2\!+\!(k\!-\!1),\!2\!+\!(j\!-\!3)}\!-  \right.\nonumber\\
&\left.\!-\varphi_k^2\eta_3\ket{3\!+\!(k\!-\!2),\!j\!-\!3}\!-\!\eta_2\ket{3\!+\!(k\!-\!2),\!1\!+\!(j\!-\!2)}\!+  \right.\nonumber\\
&\left.\!+\eta_3\ket{3\!+\!(k\!-\!2),\!2\!+\!(j\!-\!3)} \right]  \; ,
\end{align}
\begin{align}
c&^\dagger_{d \uparrow} c^\dagger_{1\uparrow}\ket{k,j}=\varphi_k^3 \left[\eta_1\ket{k\!+\!1\!,j\!+\!1}\!-\!\eta_2\ket{k\!+\!1,\!2\!+\!(j\!-\!1)}\!-\right.\nonumber\\
&\left.\!-\varphi^2_k\eta_3\ket{k\!+\!1,\!2\!+(\!j\!-\!3)}\!-\!\eta_1\ket{2\!+\!(k\!-\!1),\!j\!+\!1}\!+\right.\nonumber\\
&\left.\!+\eta_2\ket{2\!+\!(k\!-\!1),\!2\!+\!(j\!-\!1)}\!+\!\varphi^2_k\eta_3\ket{2\!+\!(k\!-\!1),\!2\!+(\!j\!-\!3)}\!-\right.\nonumber\\
&\left.\!-\eta_1\ket{2\!+\!(k\!-\!3),\!j\!+\!1}\!+\!\eta_2\ket{2\!+\!(k\!-\!3),\!2\!+\!(j\!-\!1)}\!+\right.\nonumber\\
&\left.\!+\varphi^2_k\eta_3\ket{2\!+\!(k\!-\!3),\!2\!+\!(j\!-\!3)} \right]  \; ,
\end{align}
\small
\begin{align}
c&^\dagger_{d \downarrow} c^\dagger_{1\downarrow}\ket{k,j}=\varphi_k^3 \left[\varphi_k^2\eta_3\ket{k\!+\!3,\!j\!+\!3}\!+\!\eta_2\ket{k\!+\!3,\!2\!+\!(j\!-\!1)}\!-\right.\nonumber\\
&\left.\!-\eta_3\ket{k\!+\!3,\!3\!+\!(j\!-\!2)}\!+\!\varphi_k^2\eta_3\ket{2\!+\!(k\!-\!1),\!j\!+\!3}\!+\right.\nonumber\\
&\left.\!+\eta_2\ket{2\!+\!(k\!-\!1),\!2\!+\!(j\!-\!1)}\!-\!\eta_3\ket{2\!+\!(k\!-\!1),\!3\!+\!(j\!-\!2)} \!-\right.\nonumber\\
&\left.\!-\varphi_k^2\eta_3\ket{3\!+\!(k\!-\!2),\!j\!+\!3}\!-\!\eta_2\ket{3\!+\!(k\!-\!2),\!2\!+\!(j\!-\!1)}\!+\right.\nonumber\\
&\left.\!+\eta_3\ket{3\!+\!(k\!-\!2),\!3\!+\!(j\!-\!2)} \right]  \; .
\end{align}
\normalsize

We can also derive the diagonal terms in $H^{(0)}$ involving dot operators, which we write schematically as  
\begin{align}
(n_{d\uparrow}+n_{d\downarrow})\ket{k,j}&=\ket{1\!+\!(k\!-\!1),\!j}\!+\!\ket{2\!+\!(k\!-\!2)\!,j}\!-\nonumber\\
&-\ket{3\!+\!(k\!-\!3),\!j}  \; ,
\end{align}
\begin{gather}
n_{d \uparrow} n_{d\downarrow}\ket{k,j}=\ket{2\!+\!(k\!-\!2)\!,j}\!-\!\ket{3\!+\!(k\!-\!3)\!,j}  \; .
\end{gather}

The corrected ground state $\{\ket{g^{(1)}}\}$ are the eigenvectors associated with the four-lowest eigenvalues of $H^{(0)}+H^{(1)}$ in the $\{ \ket{k,j} \}$ FPF basis. These  $|g^{(1)} \rangle$ states are divided in two doublets, with energy splitting less than $t_d/t$. Each doublet are composed of two dot FPF states (either $\ket{0}_d, c^\dagger_{d \uparrow} \ket{0}_d$ or $i c^\dagger_{d \uparrow} c^\dagger_{d \downarrow} \ket{0}_d ,-i c^\dagger_\downarrow \ket{0}_d$) together with a sum of all states in the chain.

The resulting corrected states are then used in Eq.~\eqref{eq:rho} to obtain an approximation for the dot LDOS $\tilde{\rho}_d(0)$, where we sum over the doublets with lowest energy. In general, this means we sum over only one doublet. Nonetheless, as shown in Fig.~\ref{fig:FirstOrder}, the total LDOS obtained by the approximation nicely matches the one calculated from DMRG. This is valid even when the dot's interaction is large, showing the approximation' stability. The main artifact of the approximation is that, due to the doublet splitting,  it yields a spin-polarized LDOS, while DMRG gives the correct unpolarized LDOS.

The origin of the artifact is illustrated in Fig.~ \ref{fig:spectralWeight}, which shows the components $A_k \equiv \sum_j |\langle k,j | g^{(1)} \rangle|^2$ of each state inside the ground state doublet as a function of $\epsilon_d$. For $\epsilon_d<-U_d/2$ the doublet with non-zero spectral weights is spin down polarized while for $\epsilon_d>U_d/2$ the spin up polarization prevails.

Interestingly, Fig.~ \ref{fig:spectralWeight} shows that the components of states inside the doublet are equal precisely at $\epsilon_d=0$ and $\epsilon_d=-U_d$. At these points, the state in the dot corresponds to a parafermionic mode fully localized at the quantum dot. Moving away from those points, the parafermion becomes split between dot and chain, that translates into an imbalance of spectral weights.

%\bibliography{Z4QD}
 
%

\end{document}